# Transient and steady-state dislocation creep of olivine controlled by dislocation interactions at the isostress endmember

David Wallis,[1]* Thomas Breithaupt,[1] Taco Broerse[2]

[1]Department of Earth Sciences, University of Cambridge, Downing Street, Cambridge, CB2 3EQ, UK.

[2]Department of Geoscience and Remote Sensing, Delft University of Technology, Delft, 2628 CN, The Netherlands.

*dw584@cam.ac.uk

**Abstract**

The rheological behaviour of olivine deforming by dislocation creep controls geodynamic processes that involve steady-state flow or transient viscosity evolution. Longstanding rheological models applied to both contexts assume that dislocation creep of olivine aggregates occurs close to the isostrain endmember with each grain deforming to the same strain but supporting different stress. Here, we test this assumption by constructing isostrain and isostress models based on flow laws for single crystals and comparing them to rheological data for aggregates. This analysis reveals that strain rates measured on olivine aggregates agree with those predicted by the isostress model but are an order of magnitude faster than those predicted by the isostrain model. When extrapolated to conditions typical of the shallow upper mantle, the isostress model predicts steady-state viscosities that are one to three orders of magnitude less than those predicted by the isostrain model. Furthermore, deformation close to the isostress endmember implies that transient creep occurs predominantly by dislocation interactions, suggesting viscosity changes that are approximately one order of magnitude greater than those predicted previously based on grain interactions associated with the isostrain model.

1. **Introduction**

The viscosity of Earth's upper mantle exerts a key control on the rates and styles of the majority of large-scale tectonic and geodynamic processes, including long-term convection and plate motion (Orowan, 1964; Tackley, 2000), short-term fault behaviour (e.g., Freed, 2005), and glacial isostatic adjustment (e.g., Simon et al., 2022). Therefore, constraining the viscosity of the mantle and its evolution with strain is a major outstanding objective. The most direct estimates of the viscosity of Earth's upper mantle are derived from geodetic analyses of deformation induced by rapid changes in stress, such as those imposed by major earthquakes. Typically, geodetic measurements of surface displacements and gravity changes in the aftermath of an earthquake are used to constrain models of the spatiotemporal distribution of viscosity in the upper mantle (e.g., Hoechner et al., 2011; Freed et al., 2012; Broerse et al., 2015; Masuti et al., 2016; Muto et al., 2019). Notably, several such analyses indicate that viscosity depends nonlinearly on the applied stress, implying the activity of a dislocation-mediated deformation mechanism (e.g., Freed et al., 2006, 2012; Masuti et al., 2016). Such models also indicate that viscosity increases with time and strain after a stress change in a manner that cannot be accounted for using flow laws calibrated only for steady-state deformation (Freed et al., 2010, 2012; Masuti et al., 2016; Muto et al., 2019). This observation implies that the mantle undergoes transient creep, in which viscosity evolves during deformation due to changes in microstructure and micromechanical state, as is widely observed in laboratory experiments following changes in stress (e.g., Post, 1977; Smith and Carpenter, 1987; Chopra, 1997; Hanson and Spetzler, 1994; Hansen et al., 2021).

The accuracy of inferred mantle viscosities depends on the use of an appropriate rheological model for dislocation creep of olivine, which constitutes > 60% of the upper mantle and therefore exerts the dominant control on its rheological properties (Zimmerman and Kohlstedt, 2004). As olivine exhibits pronounced elastic (Abramson et al., 1997) and viscous (Bai et al., 1991) anisotropy, modelling the deformation of an olivine aggregate is not straightforward. At pressures sufficiently high to prevent the opening of voids, successful models must maintain a compatible strain field among mechanically anisotropic grains. A compatible strain field could occur through either homogeneous or heterogeneous strain and these two possibilities lead to two endmember models, that is an isostrain endmember (Taylor, 1938) and an isostress endmember (Sachs, 1928). At the isostrain endmember, each grain deforms to the same strain. For viscous flow in three dimensions, this model imposes the constraint that each grain has the same strain and strain rate tensors. If the grains are mechanically anisotropic, the isostrain model necessitates that grains with different lattice orientations support different stresses. In contrast, at the isostress endmember, each grain is subjected to the same stress tensor applied to the bulk material. In this case, mechanical anisotropy dictates that grains with different lattice orientations deform to different strains at different strain rates. Strain compatibility between grains can then be facilitated by intragranular strain heterogeneity, whereby spatial gradients in plastic strain arise in grain interiors (Paterson, 1969; Ashby, 1970). These intragranular strain gradients are maintained by the presence of geometrically necessary dislocations (Fleck et al., 1994). The isostrain and isostress endmember models make significantly different predictions about the nature of transient creep and steady-state flow.

The isostrain model was first considered in detail for Earth materials in relation to early experiments on water ice (Ashby and Duval, 1985). In the isostrain model, transient creep occurs due to grain interactions that transfer stress between two sets of grains with different crystallographic orientations (Karato, 1998, 2021; Masuti and Barbot, 2021). In anisotropic crystal structures, such as ice and olivine, the slip systems upon which dislocations glide can have viscosities that differ by orders of magnitude (Duval et al., 1983; Bai et al., 1991). At the beginning of deformation, dislocation glide is assumed to occur at the fastest rate in grains with orientations that result in high resolved shear stress on the weakest slip system. However, to maintain strain compatibility among adjacent grains, this initial deformation progressively transfers stress to grains in which the weakest slip system has low resolved shear stress. This transfer of stress is assumed to activate stronger slip systems in the latter set of grains, with the strain rate on the strongest slip system controlling the steady-state strain rate of the aggregate. The transition from the bulk strain rate initially being controlled by the weak slip system to later being controlled by the strong slip system results in a strain-hardening transient. As the steady-state strain rate at this endmember is limited by the strain rate on the strongest slip system, this model provides an upper bound on the strength of the aggregate.

In contrast, in the isostress model each grain behaves as a single crystal subjected to the same stress tensor. In this case, the steady-state strain rate of an aggregate is typically dominated by deformation on the easy slip system such that the isostress model predicts an aggregate to be weaker than predicted by the isostrain model. Furthermore, in the absence of intergranular stress transfer, any transient creep must arise from intragranular processes. In the case of olivine, the dominant intragranular process appears to be elastic interactions among dislocations (Wallis et al., 2017, 2021, 2022; Hansen et al., 2021). This inference is based on recent deformation experiments and microstructural observations of olivine single crystals deformed at room temperature (Hansen et al., 2019) and temperatures in the range 1200–1300°C (Wallis et al., 2017, 2021; Hansen et al., 2021). Following reductions in stress, single crystals of olivine undergo time-dependent reverse strains (Hansen et al., 2021). This reverse deformation reveals the existence of a back stress within the material that counteracts the applied stress and is inferred to arise from elastic interactions among dislocations (Hansen et al., 2021). As dislocation density changes during deformation, so does the intensity of the elastic interactions among dislocations that resist dislocation motion (Hansen et al., 2021; Wallis et al., 2021). The increasing elastic interactions and associated back stress counteract the applied stress, leading to a strain-hardening transient (Hansen et al., 2021). We note that dislocations impart significant intragranular stress heterogeneity to both single crystals (Wallis et al., 2017; Hansen et al., 2021) and aggregates (Wallis et al., 2021) of olivine, which causes elastic interactions that generate back stress. However, to be clear, the term 'isostress' refers to a lack of systematic differences in grain-average stress supported by grains with different lattice orientations regardless of intragranular stress heterogeneity over length scales smaller than the grain size.

Deformation of olivine aggregates has long been considered to occur close to the isostrain endmember (Karato, 1998, 2021; Hansen et al., 2016; Masuti et al., 2019, 2021). This inference is based primarily on the observation that steady-state strain rates measured on aggregates lie among the strain rates measured on single crystals oriented to activate the strong slip systems (Chopra and Paterson, 1984; Mei and Kohlstedt, 2000; Keefner et al., 2011). This observation has been taken to indicate that the strong slip systems have been activated in the aggregates and become rate limiting due to grain interactions (Karato, 1998, 2021). This analysis is often justified on the basis of the von Mises criterion, which states that five independent slip systems are required to allow a crystal to deform to an arbitrary strain by dislocation glide (von Mises, 1913). However, olivine lacks five independent slip systems and therefore strain-producing mechanisms other than dislocation glide must also operate in the isostrain model (Paterson, 1969). Despite the operation of these alternative mechanisms, it is likely that the strongest of the available slip systems still limits the bulk strain rate under isostrain conditions depending on the relative efficiency of each mechanism. The isostrain interpretation has provided the conceptual basis for recent geodetic analyses employing nonlinear adaptations of the Burgers model, in which the two viscous elements represent the weak slip system, which controls the initial transient strain rate, and the strong slip system, which controls the strain rate at steady state (Masuti et al., 2016; Muto et al., 2019; Masuti and Barbot, 2021). Such formulations are typically parameterised using flow laws for individual slip systems calibrated at steady state (Masuti et al., 2016; Muto et al., 2019).

Here, we query the relevance of all three of these aspects of the isostrain interpretation. First, we contend that previous face-value comparisons between strain rates measured on aggregates and those measured on single crystals are invalid because the resolved shear stresses on each slip system (i.e., the shear stress acting on the slip plane in the slip direction) averaged over all grains in an aggregate are less than those in experiments on single crystals that are oriented to maximise the resolved shear stresses acting on particular slip systems. That is, even if grains do not interact, aggregates will deform at slower strain rates than single crystals oriented to activate the easy slip system due simply to differences in resolved shear stress. Therefore, we pose the null hypothesis that, once the resolved shear stresses acting on each slip system are properly accounted for, steady-state dislocation creep of coarse-grained aggregates of olivine occurs at the strain rates expected of an ensemble of grains that do not interact by intergranular stress transfer, i.e., at the strain rates predicted by the isostress model. Second, we note that the Von Mises criterion in a strict sense neglects the potential for intragranular strain heterogeneity or strain-producing processes other than dislocation glide (Paterson, 1969), both of which are ubiquitous in deforming olivine aggregates and potentially reduce the impact of the strong slip system on the bulk strain rate. Lastly, we emphasise that single crystals with no neighbouring grains exhibit marked transient creep (Kohlstedt and Goetze, 1974; Durham and Goetze, 1977; Durham et al., 1979; Hanson and Spetzler, 1994; Cooper et al., 2016; Hansen et al., 2021), which must arise from intragranular processes. As the same intragranular processes must also operate within grains in aggregates, intergranular stress transfer in the isostrain model is at best an incomplete description of transient creep. Thus, models of transient creep based on steady-state flow laws for the soft and hard slip systems (Masuti et al., 2016; Muto et al., 2019; Masuti and Barbot, 2021) are missing an important set of intragranular physical processes.

To test our null hypothesis, we compare predicted strain rates from isostress and isostrain models based on flow laws for each slip system measured on single crystals to strain rates measured on aggregates at steady state. Furthermore, we compare strain rates measured on aggregates at the start of deformation prior to any hardening to a recently calibrated flow law that can predict strain rates in the absence of intragranular hardening. This comparison allows us to explore the extent to which either steady-state strain rates are controlled by the strong slip system and transients arise from grain interactions (i.e., the isostrain model) or steady-state strain rates are dominated by the weak slip system and transients arise from intragranular processes (i.e., the isostress model). We then explore the implications of discriminating between the isostrain and isostress models for deformation in the steady-state and transient regimes under natural conditions.

## 2. Methods

*2.1. Experimental materials*

To compare the strain rates of aggregates of olivine to those expected from combinations of individual slip systems, the experimental datasets must be carefully selected to properly control for the full range of variables that can influence steady-state strain rates. These variables include composition, grain size, temperature, pressure, oxygen fugacity, the activities of $H_2O$ and $SiO_2$, and the presence or absence of partial melt (Bai et al., 1991; Keefner et al., 2011; Warren and Hansen, 2023). For aggregates of olivine, we work primarily with the rheological data measured in the experiments of Keefner et al. (2011) on Åheim dunite, a rock consisting of approximately 95% olivine. These experiments are particularly appropriate because the coarse grain size of 0.9 mm of the dunite prevents significant contributions from deformation mechanisms other than dislocation creep and because these experiments were conducted under carefully controlled oxygen fugacities. These samples were predried, had the activity of silica buffered by the presence of small amounts of orthopyroxene, and lacked microstructural evidence for the presence of partial melt. We also use data from samples 4392, 4462, and 4526 from selected experiments by Chopra and Paterson (1981, 1984) and Doukhan et al. (1984) that were deformed at temperatures and oxygen fugacities that are useful in making comparisons to the data of Keefner et al. (2011) and the flow law for dislocation glide in the absence of hardening (Breithaupt et al., 2023). These samples also consisted of either Åheim or Anita Bay dunite that was predried and in which the silica activity was again buffered by orthopyroxene. However, in these experiments the oxygen fugacity is less well constrained than in the experiments of Keefner et al. (2011). Sample 4392 may have retained residual hydrous defects (Chopra and Paterson, 1984), and sample 4526 displayed glassy films on grain boundaries that suggest the presence of small fractions of partial melt (Doukhan et al., 1984). These factors provide potential explanations for why the steady-state strain rates in these experiments are slightly higher than those in the experiments of Keefner et al. (2011).

To represent the strain rates on individual slip systems, we use flow laws from two previous studies. To calculate steady-state strain rates, we use the flow laws of Bai et al. (1991). These flow laws were derived from uniaxial experiments performed on single crystals of dry San Carlos olivine with the oxygen fugacity carefully controlled during the experiments. We use the flow laws from experiments in which the silica activity was set by the presence of orthopyroxene on the surface of the sample. These factors make the flow laws of Bai et al. (1991) the most suitable upon which to formulate comparisons to the data from olivine aggregates by Keefner et al. (2011) for the purpose of steady-state flow. Notably, as these flow laws were calibrated at steady state after the initial transients, they intrinsically include the effects of intragranular hardening by dislocation interactions. Moreover, having been calibrated on single crystals, these flow laws naturally omit intergranular hardening by stress transfer among grains. In contrast, to calculate strain rates at the onset of deformation before any hardening, we use the flow law of Breithaupt et al. (2023) for dislocation glide at high temperatures. This flow law was calibrated on data collected at steady state in experiments performed on aggregates by Hansen et al. (2011) and Keefner et al. (2011), as part of a wider inversion for both dislocation-glide and dislocation-recovery parameters. The silica activity in these experiments was buffered by the presence of small amounts of orthopyroxene. Only experiments in which the oxygen fugacity was controlled at the nickel-nickel oxide buffer were used in the calibration. The flow law has been demonstrated to well predict transient creep of single crystals deformed under oxygen fugacities controlled at the nickel-nickel oxide buffer in stress-reduction experiments (Hansen et al., 2021). Variation in oxygen fugacity has not yet been demonstrated to impact transient strain rates limited by dislocation glide, and thus, we do not correct for variation in oxygen fugacity when comparing the flow law for dislocation glide calibrated at the nickel-nickel oxide buffer to transient strain rates measured at the iron-wüstite buffer. Instead, this comparison provides an initial test for any effect of oxygen fugacity on rates of dislocation glide.

The remaining variables to consider across the sample sets are composition and pressure. San Carlos olivine (Jin et al., 1994), Åheim dunite (Chopra and Paterson, 1981; Keefner et al., 2011), and Anita Bay dunite (Chopra and Paterson, 1981) all have Mg/(Mg+Fe) in the range 0.89–0.93 and are therefore similar in major-element chemistry. Differences in major- and trace-element chemistry of the magnitudes present between, for example San Carlos and Åheim olivine, do not generate significant differences in creep rates and dependencies under the conditions of interest (Jin et al., 1994). The experiments on single crystals were conducted at a confining pressure of 0.1 MPa (Bai et al., 1991) whereas those on dunites were conducted at a confining pressure of 300 MPa (Chopra and Paterson, 1981, 1984; Doukhan et al., 1984; Keefner et al., 2011). The effect of pressure (Dixon and Durham, 2018) on dislocation creep of aggregates of dry olivine over this pressure difference is only factor of approximately 1.4 in strain rate, which we account for in the models as described below.

*2.2. Electron backscatter diffraction*

We measured the crystallographic preferred orientation (CPO) of Åheim dunite, which was used as starting material for the experiments of Keefner et al. (2011), on a piece supplied by the Rock and Mineral Physics Laboratory, University of Minnesota. The sample was cut parallel to the foliation, polished with progressively finer diamond grits, and finished with colloidal silica. We acquired orientation measurements by electron backscatter diffraction (EBSD) on an FEI Quanta 650 field emission gun environmental scanning electron microscope in the Department of Earth Sciences, University of Oxford, equipped with an Oxford Instruments AZtec acquisition system and Nordlys Nano EBSD detector. The map consists of 808 × 346 data points at a step size of 10 μm. Misindexed pixels that differed in orientation from their neighbours by > 10° were removed and unindexed pixels surrounded by > 6 neighbours within the same grain were iteratively assigned the average orientation of their neighbours. The resulting dataset consists of the average orientation of each of the resulting 418 grains. Figure 1 presents the CPO of our sample, which is near identical to that plotted in Figure 8 of Keefner et al. (2011), providing confidence that the CPO of the Åheim dunite is homogeneous and therefore that the measured CPO is representative of that used in the deformation experiments of Keefner et al. (2011). We used the CPO of the undeformed

Åheim dunite that was used as starting material for the experiments (Keefner et al., 2011) in our calculations for the isostress and isostrain models. We modelled compression of the dunite with the maximum principal stress in the same direction relative to the CPO as was applied in the experiments of Keefner et al. (2011).

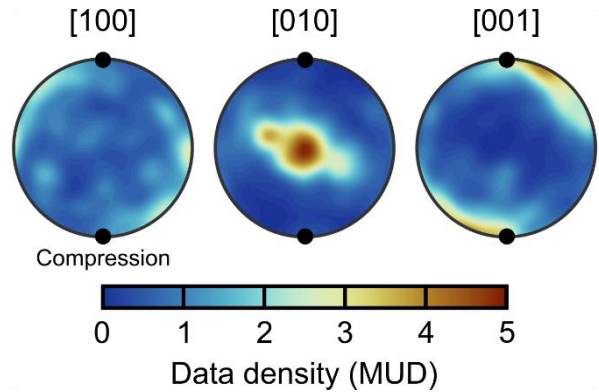

**Figure 1.** The CPO of Åheim dunite used in the experiments of Keefner et al. (2011) and as input to the isostress and isostrain models. Black dots indicate the direction of maximum principle stress applied in the experiments and models. The colour scale is in multiples of uniform distribution (MUD).

*2.3. Models of steady-state deformation*

*2.3.1. Isostress model for steady-state deformation*

To model the isostress endmember strain rate of an olivine aggregate, we utilise the approach of Tielke et al. (2016), which is a simple formulation of the classical isostress model of Sachs (1928). In this conceptual end-member model for the deformation of an aggregate of grains, each grain is assumed to be subjected to the same local stress tensor, which is the macroscopic stress tensor applied to the bulk material. Therefore, this approach does not model stress heterogeneity arising from transfer of stress between grains. We note that in detail stress heterogeneity is present within deformed olivine aggregates due to the stress fields of dislocations (Wallis et al., 2017, 2020; 2021; Hansen et al., 2021). However, these stresses are also present in deformed single crystals (Wallis et al., 2017, 2020) and result from a different set of physical processes compared to intergranular stress transfer. Specifically, the stress heterogeneity imparted by dislocations controls intragranular hardening by long-range dislocation interactions (Wallis et al., 2017, 2020; 2021; Hansen et al., 2021), the presence or absence of which we account for in the flow laws input to the model (Bai et al., 1991; Breithaupt et al., 2023) as described below. Therefore, in using the term 'isostress model', we refer specifically to a lack of systematic differences in the average stress tensor supported by subpopulations of grains with different crystallographic orientations. The benefit of this minimalist model is that it can test the extent to which strain rates measured on olivine aggregates can be explained simply by resolving the stress tensor applied to the bulk rock onto the slip systems in each grain in the absence of other confounding physical processes.

The modelling procedure is based on the computation of stresses and strain rates in both the sample and crystallographic reference frames (Tielke et al., 2016). We define the deviatoric stress tensor, **σ**, in the sample reference frame, making it consistent with the loading axis applied experimentally to the Åheim dunite and the grain orientations within. The deviatoric stress tensor is rotated into the crystallographic reference frame (indicated by the prime symbol) of each grain by

$$\boldsymbol{\sigma}' = \mathbf{R}\boldsymbol{\sigma}\mathbf{R}^{\mathrm{T}}, \tag{1}$$

where **R** is the passive rotation matrix defining the rotation from the crystal reference frame to the sample reference frame. Slip systems, $\alpha$, are defined by their unit Burgers vector, **b**, and unit slip-plane normal, **n**. We consider the (010)[100], (001)[100], (100)[001], and (010)[001] slip systems (19). The resolved shear stress on each slip system, $\tau^\alpha$, is calculated by

$$\tau^\alpha = \boldsymbol{\mu}^\alpha : \boldsymbol{\sigma}', \tag{2}$$

where **μ**$^\alpha$ is the symmetric Schmid tensor determined as

$$\boldsymbol{\mu}^\alpha = \frac{1}{2}\left(\mathbf{b}^\alpha \otimes \mathbf{n}^\alpha + \mathbf{n}^\alpha \otimes \mathbf{b}^\alpha\right), \tag{3}$$

where $\otimes$ is the tensor outer product. The corresponding differential stress, $\sigma^\alpha$, acting on each slip system is calculated as

$$\sigma^\alpha = |2\tau^\alpha|. \tag{4}$$

Then, we determine the strain rate, $\dot{\varepsilon}^\alpha$, due to dislocation motion on each slip system using the flow laws for single crystals of San Carlos olivine. To calculate strain rates during steady-state flow, inherently capturing the effects of intragranular hardening during the preceding transient, we use the flow laws of Bai et al. (1991), following the general form

$$\dot{\varepsilon}^\alpha = \dot{\varepsilon}^\alpha\left(\sigma^\alpha, T, P, f_{O_2}\right)\mathrm{sgn}\left(\tau^\alpha\right), \tag{5}$$

where $f_{O_2}$ is oxygen fugacity. We use the three flow laws for single crystals of olivine with the silica activity buffered by orthopyroxene and that were deformed in different orientations chosen to activate the slip systems listed above. As the (001)[100] and (100)[001] slip systems are activated simultaneously in experiments on a single crystal orientation and always have the same Schmid factor, we calculate their combined strain rate using a single flow law (Bai et al., 1991). Here, the effect of pressure, $P$, is added to the original flow laws by incorporating more recent estimates of the activation volume, $\Delta V$, for single crystals deformed in each orientation (Raterron et al., 2012). Specifically, the slip-system specific activation energies are adjusted with the relevant activation volume as

$$Q(P) = Q_{\text{ref}} + (P - P_{\text{ref}})\Delta V, \tag{6}$$

where $Q_{\text{ref}}$ is the fitted activation energy at the reference pressure $P_{\text{ref}} = 0.1$ MPa, which corresponds to the pressures at which the flow laws of Bai et al. (1991) were calibrated. This pressure effect is small in that, for example, it reduces strain rates predicted by the isostress model at a pressure of 300 MPa by approximately 30% relative to those predicted at a pressure of 0.1 MPa. The corresponding shear strain rates, $\dot{\gamma}^\alpha$, are

$$\dot{\gamma}^\alpha = 2\dot{\varepsilon}^\alpha, \tag{7}$$

which can be summed across slip systems to give the strain rate tensor of each grain in the crystallographic reference frame as

$$\dot{\varepsilon}' = \sum_{\alpha=1}^{3} \boldsymbol{\mu}^\alpha \dot{\gamma}^\alpha. \tag{8}$$

This strain-rate tensor is rotated into the sample reference frame, giving $\dot{\varepsilon}_{\text{grain}}$, by

$$\dot{\varepsilon}_{\text{grain}} = \mathbf{R}^{\text{T}} \dot{\varepsilon}' \mathbf{R}. \tag{9}$$

Assuming as a simple approximation that each grain has the same volume, we estimate the strain rate of the aggregate by taking the arithmetic mean of $\dot{\varepsilon}_{\text{grain}}$ across all grains. Finally, we take the component of the strain rate of the aggregate that is parallel to the applied loading axis.

*2.3.2. Isostrain model for steady-state deformation*

We contrast our calculations of the isostress endmember with calculations assuming isostrain, the alternative conceptual end-member model for the deformation of aggregates (Taylor, 1938). In the isostrain limit, each grain is subjected to the same macroscopic strain tensor. Due to plastic anisotropy, each grain possesses its own local stress tensor. The macroscopic stress tensor then corresponds to an average of all of the grain-specific local stress tensors. Application of the isostrain model to olivine is challenging because olivine does not possess sufficient independent slip systems to produce an arbitrary strain tensor (Paterson, 1969). Previous studies have addressed this issue by incorporating stronger fictitious slip systems (Tommasi et al., 2000) or by partially relaxing the isostrain assumption to require the grain-specific strain tensor to closely match, rather than exactly match, the macroscopic strain tensor (Hansen et al., 2016). If stronger fictitious slip systems are introduced, the modelled aggregate behavior may be strengthened by an arbitrary amount depending on the assumed strength of the fictitious slip system (Castelnau et al., 2008). Alternatively, if the isostrain assumption is partially relaxed, alternative unmodelled deformation processes are implicitly assumed to operate that would allow each grain to completely fulfil the isostrain condition. These unmodelled deformation processes are assumed to have negligible strength. In this work, we instead take inspiration from published isostrain models of transient creep that neglect the tensorial aspects of deformation and instead assume a scalar formulation of isostrain (Ashby and Duval, 1985; Karato, 1998, 2021). We focus on the component of strain in the compression direction and enforce the isostrain condition on this component only. We note that enforcing the isostrain condition on all components will likely strengthen the modelled aggregate behaviour, and thus, the predicted strain rates provide upper bounds on those that would be predicted by complete isostrain approaches.

We define the grain-specific, one-dimensional flow law to be

$$\dot{\varepsilon}_i = A_i \sigma_i^n, \tag{10}$$

where the subscript indicates that a property is specific to the $i^{\text{th}}$ grain, $n$ is the stress exponent taken to be equal to 3.5 for all grains (Bai et al., 1991), and $A_i$ is a grain-specific flow-law parameter. That $A_i$ is variable among grains reflects that each grain has a different crystallographic orientation.

In the isostrain limit, at steady-state, each grain experiences the bulk plastic strain rate $\dot{\varepsilon}_{\text{isostrain}}$. The grain-specific stress is therefore given by

$$\sigma_i = \dot{\varepsilon}_{\text{isostrain}}^{1/n} A_i^{-1/n}. \tag{11}$$

In the alternative isostress limit, all grains are subjected to the macroscopic stress $\sigma_i = \sigma$. Thus, we can determine the grain-specific flow-law constants by

$$A_i = \dot{\varepsilon}_{i,\text{isostress}} \sigma^{-n}, \tag{12}$$

where $\dot{\varepsilon}_{i,\text{isostress}}$ are the grain-specific strain rates in the compression direction determined using the isostress methodology detailed above.

The macroscopic stress is determined by the average of the grain-specific stresses. Taking the average of Equation 12, and substituting Equation 11, we find

$$\sigma = \dot{\varepsilon}_{\text{isostrain}}^{1/n} \frac{1}{N} \sum_{i=1}^{N} \sigma \dot{\varepsilon}_{i,\text{isostress}}^{-1/n}, \tag{13}$$

where $N$ is the total number of grains. Here, we make the assumption again that each grain has the same volume. We then divide by the macroscopic stress and rearrange to find the steady-state plastic strain-rate under the isostrain model

$$\dot{\varepsilon}_{\text{isostrain}} = \left( \frac{1}{N} \sum_{i=1}^{N} \dot{\varepsilon}_{i,\text{isostress}}^{-1/n} \right)^{-n}. \tag{14}$$

*2.4. Models of viscosity evolution*

To model viscosity evolution during transient creep, we employ the models Karato (2021) and Breithaupt et al. (2023), which are based on grain interactions and dislocation interactions respectively. Breithaupt et al. (2023) provide a model for the temporal evolution of strain rate resulting from changes in the density of dislocations and the strength of interactions among them. In contrast, Karato (1998, 2021) proposed a model for the evolution of strain rate based on transfer of stress from the weakest slip system to a stronger slip system due to grain interactions. We perform a simple comparison of the behaviours of these models to demonstrate that they differ significantly for applications to geodynamic questions.

*2.4.1. Model of intergranular hardening by grain interactions*

The model of intergranular hardening by grain interactions proposed by Karato (2021) assumes a coupled system of a weak and a strong slip system. Following a stress increase, the initial increment of deformation occurs on the weak slip system in grains with orientations that provide this slip system with high resolved shear stress. Progressive deformation transfers stress from this weak slip system to stronger slip systems that are activated to maintain a homogeneous strain field without strain incompatibilities at grain boundaries. Consequently, the initial increment of deformation is controlled by the viscosity of the weak slip system whereas eventual steady-state flow is controlled by the viscosity of the strong slip system. Different from the original approach of Karato (2021), we assume a pre-stressed material by applying a background stress (in the same direction as the stress increase) that results in an existing transfer of stress up to a converged state that is present prior to the application of a new constant stress. The model assumes steady-state deformation of each slip system so that the viscosities of individual slip systems do not evolve with strain or time. We use the pressure-adjusted flow laws for individual slip systems from Bai et al. (1991), namely the weak (010)[100] and strong (010)[001] slip systems.

In the model of grain interactions (Karato, 2021), strain rate depends of the difference between the applied differential stress, $\sigma$, and the transferred stress, $S$, which defines effective stresses, $\sigma_1$ and $\sigma_2$ acting on the weak and strong slip systems, respectively, as

$$\sigma_1 = \sigma - S_1 \text{ and } \sigma_2 = \sigma - S_2, \tag{15}$$

with the condition that

$$S_1 = -S_2 \tag{16}$$

to achieve a net transferred stress of zero. The strain rate for each slip system then becomes a function of the effective stresses, $\sigma_1$ and $\sigma_2$, of the form

$$\dot{\varepsilon}_{1,2} = f\left(\sigma_{1,2}, A, f_{O_2}^m, \exp\left(\frac{-Q}{RT}\right)\right), \tag{17}$$

with pre-exponent $A$, stress exponent $n$, oxygen fugacity $f_{O_2}$ with exponent $m$, activation energy $Q$, ideal gas constant $R$, and absolute temperature $T$.

The constitutive equations from Bai et al. (1991) incorporate multiple mechanisms that are active concurrently or sequentially. Each of these mechanisms separately follow a power-law flow law of the form

$$\dot{\varepsilon} = A\sigma|\sigma^{n-1}|f_{O_2}^m \exp\left(\frac{-Q}{RT}\right). \tag{18}$$

For the weak slip system, (010)[100], the constitutive equation includes 3 mechanisms, with mechanism b and c operating sequentially, while this combination acts concurrently with mechanism a as

$$\dot{\varepsilon}_1 = \dot{\varepsilon}_a + (\dot{\varepsilon}_b^{-1} + \dot{\varepsilon}_c^{-1})^{-1}. \tag{19}$$

The strong slip system, (010)[001], consists of two concurrent mechanisms, d and e, which lead to the constitutive equation

$$\dot{\varepsilon}_2 = \dot{\varepsilon}_d + \dot{\varepsilon}_e. \tag{20}$$

We compute oxygen fugacity, $f_{O_2}$, as a function of temperature and pressure at the FMQ buffer (O'Neill, 1987a, 1987b; Ballhaus et al., 1991; Frost, 1991). Table S1, given in Supporting Information, contains the values of the parameters for each mechanism that we use in the model of grain interactions.

The resulting strain rate of the coupled system is (Equation 14 of Karato (2021))

$$\dot{\varepsilon} = \frac{1}{2}(\dot{\varepsilon}_1 + \dot{\varepsilon}_2). \tag{21}$$

To maintain homogeneous strain, the time derivative of transferred stress is (Equation 17 of Karato (2021))

$$\dot{S}_1 = \frac{\mu}{2}(\dot{\varepsilon}_1 - \dot{\varepsilon}_2), \tag{22}$$

where $\mu$ is the shear modulus. We integrate the transferred stress rate, taking small time steps, and subsequently use the updated effective stress to calculate a new strain rate, until we reach steady-state strain rates.

*2.4.2. Model of intragranular hardening by dislocation interactions*

We use the model of intragranular hardening from Breithaupt et al. (2023) to model strain rates by dislocation creep as a function of the difference between the applied stress, $\sigma$, and an internal back stress, $\sigma_b$, following

$$\dot{\epsilon} = A \exp\left(\frac{-Q}{RT}\right) \sigma_\rho^2 \sinh\left(\frac{\sigma - \sigma_b}{\sigma_{\text{ref}}(T)}\right), \tag{23}$$

were $A$ is a pre-exponential constant and $Q$ is the pressure-adjusted activation energy (see Equation 6). Breithaupt et al. (2023) decompose the back stress into a Taylor stress, $\sigma_\rho$, resulting from dislocation interactions and a threshold stress, $\sigma_d$, that describes the stress necessary to expand a dislocation loop within a grain of a particular size such that

$$\sigma_b = \sigma_\rho + \sigma_d, \tag{24}$$

where the Taylor stress relates to the scalar dislocation density, $\rho$, as

$$\sigma_\rho = \alpha \mu b \sqrt{\rho}, \tag{25}$$

where $\alpha$ is a constant, $\mu$ is the shear modulus, and $b$ is the magnitude of the Burgers vector. The threshold stress is defined as

$$\sigma_d = \beta \mu b / d, \tag{26}$$

where $\beta$ is a constant and $d$ is grain size. The effective stress is scaled by a temperature-dependent reference stress, defined as

$$\sigma_{\text{ref}}(T) = \sigma_P^* \frac{RT}{Q}, \tag{27}$$

where $\sigma_P^*$ is the strength of local barriers to dislocation motion, and $Q$ is the pressure-adjusted activation energy.

In this model, transient behaviour results from the evolution of dislocation density by processes that increase (storage) or decrease (recovery) the dislocation density. Breithaupt et al. (2023) account for dynamic dislocation storage due to dislocation glide and account for static recovery facilitated by pipe diffusion and grain-boundary diffusion. As the Taylor stress is dependent on dislocation density, the effective stress changes as the dislocation density evolves. Breithaupt et al. (2023) account for the effect of changing dislocation density in the evolution of the Taylor stress (their Equation 29) by

$$\dot{\sigma}_\rho = M \left( \frac{\sigma_\rho + \sigma_d}{\sigma_\rho} \dot{\varepsilon} - \mathcal{R}_{\mathrm{gb}}(T) \sigma_\rho^3 \sigma_d - \mathcal{R}_{\mathrm{pipe}}(T) \sigma_\rho^5 \right), \tag{28}$$

where $M$ is the plastic modulus, which controls the overall rate of the transient. $\mathcal{R}_{\mathrm{gb}}(T)$ and $\mathcal{R}_{\mathrm{pipe}}(T)$ are rate coefficients for static recovery by grain-boundary diffusion and pipe diffusion, respectively, defined as

$$\mathcal{R}_{\mathrm{gb}}(T) = \mathcal{R}_{\mathrm{gb}}^* \exp\left(\frac{-Q}{\mathrm{R}T}\right) \tag{29}$$

and

$$\mathcal{R}_{\mathrm{pipe}}(T) = \mathcal{R}_{\mathrm{pipe}}^* \exp\left(\frac{-Q}{\mathrm{R}T}\right), \tag{30}$$

where $\mathcal{R}_{\mathrm{gb}}^*$ and $\mathcal{R}_{\mathrm{pipe}}^*$ are pre-exponential constants for grain-boundary diffusion and pipe diffusion, respectively, and $Q$ is the pressure-adjusted activation energy. We start with a dislocation density that conforms to the steady-state value for the initial background stress and update the Taylor stress at each time step and use this value to compute a new strain rate. The values of the parameters that we use in the model of dislocation interactions are given in Table S2 of Supporting Information.

## 3. Results

Figures 2a and 2b present comparisons of steady-state strain rates (normalised to two different temperatures and oxygen fugacities) measured from dunite by Keefner et al. (2011) to those predicted by the isostress and isostrain models. In these figures, strain rates are normalised by those predicted by the power-law flow law for dunites of Keefner et al. (2011) that was originally fit to the experimental data. This normalisation allows a straightforward comparison of the extent to which the isostress and isostrain models agree with the experimental data. The solid black curves indicate strain rates predicted by this isostress model based on the flow laws of Bai et al. (1991) for each slip system and the CPO of the dunite. These flow laws were originally calibrated at room pressure. To facilitate a more precise comparison with data collected on aggregates at a pressure of 300 MPa, we incorporated more recently calibrated pressure effects (Raterron et al., 2012). As these flow laws were calibrated on single crystals at steady state after the initial transients, they intrinsically include the effects of intragranular hardening by dislocation interactions. Strikingly, the steady-state strain rates predicted by the isostress model are in excellent agreement with the data measured on dunite across the full range of differential stresses. This agreement between the strain rates predicted by the isostress model and those measured in experiments indicates that dunites deform at the strain rates expected of an ensemble of grains that do not mechanically interact. In contrast, steady-state strain rates predicted by the isostrain model based upon the same single-crystal flow laws, indicated by the dashed black curves, are approximately an order of magnitude slower than those measured in the experiments. The low strain rates predicted by the isostrain model result from stress transfer from the easy slip system to the strong slip system. Overall, the magnitudes of steady-state strain rates measured in experiments are in good agreement with predictions of the isostress model and poor agreement with predictions of the isostrain model.

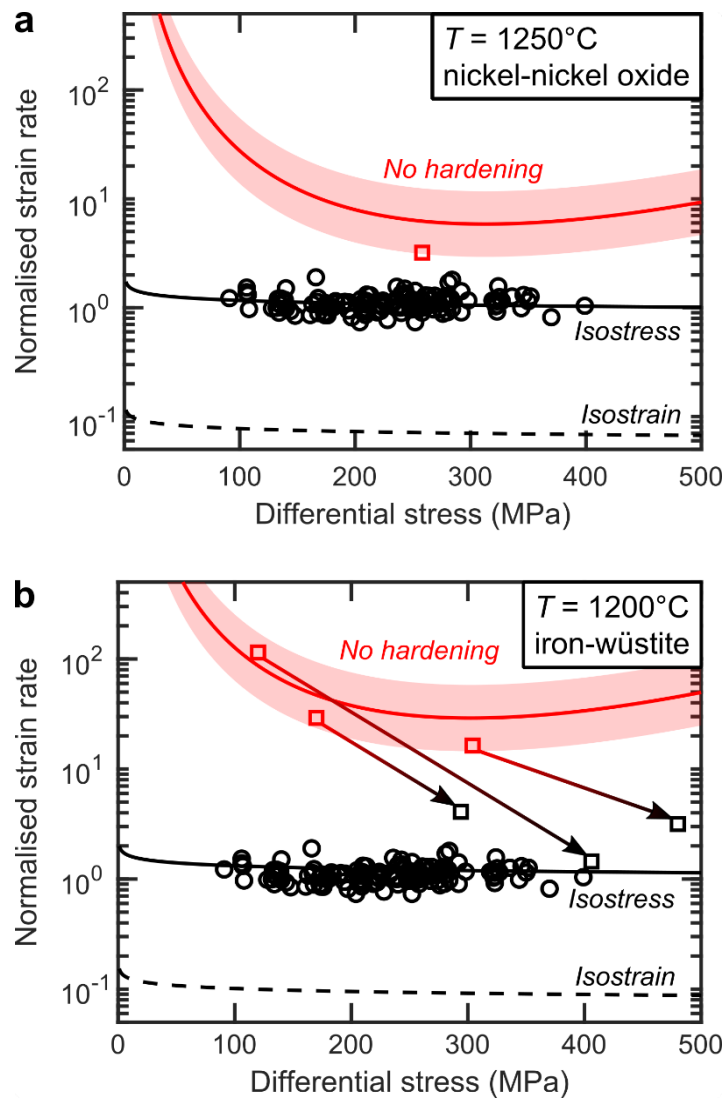

**Figure 2.** Strain rates measured in experiments and predicted by the isostress and isostrain models. **(a)** Measured and modelled strain rates normalised by division by the strain rates

predicted by the flow law measured in the experiments (Keefner et al., 2011). Black circles are strain rates measured during steady-state flow (Keefner et al., 2011) normalised to a temperature of 1250°C and oxygen fugacity of $5\times10^{-3}$ Pa at the nickel-nickel oxide buffer. The solid black curve indicates strain rates predicted by the isostress model using flow laws measured from single crystals undergoing steady-state flow, which inherently incorporates the effects of intragranular hardening (Bai et al., 1991). The dashed black curve indicates strain rates predicted by the isostrain model, also using flow laws measured from single crystals undergoing steady-state flow, which represents hardening due to elastic interactions among grains with different orientations. The red curve indicates strain rates predicted by a flow law for dislocation glide (Breithaupt et al., 2023) in the absence of hardening for an initial dislocation density of $8\times10^{11}$ m$^{-2}$, with the red shaded region indicating the effect of varying the initial dislocation density by up to a factor of 2. The red datum is a strain rate measured at the onset of deformation before hardening (Keefner et al., 2011). **(b)** The same as (a) but at a temperature of 1200°C and oxygen fugacity of $3\times10^{-6}$ Pa at the iron-wüstite buffer to allow comparison of the models to the initial (red squares) and steady-state (black squares) strain rates measured in additional experiments (Chopra and Paterson, 1981, 1984; Doukhan et al., 1984). Arrows link data from the same experiment to indicate the evolution of viscosity during the transients. Uncertainties on data points are typically less than the marker size.

In addition to analysing the magnitudes of the strain rates predicted by the isostress and isostrain models, we also examine their predicted dependencies on oxygen fugacity and temperature in Figure 3. The isostress and isostrain models predict different dependencies of strain rate on oxygen fugacity and temperature due to the different dependencies of the individual slip systems on these variables (Bai et al., 1991). The data plotted in orange and purple in Figure 3 were collected at the nickel-nickel oxide and iron-wüstite buffers respectively and the separation between them indicates the oxygen-fugacity dependence of the steady-state strain rate. The oxygen-fugacity dependence of the strain rate predicted by the isostress model, indicated by the separation between the orange and purple solid lines, is in reasonable agreement with that exhibited by the measured strain rates (Keefner et al., 2011). For comparison, the conventional power-law flow law (Keefner et al., 2011) originally fit to the experimental data is indicated by the dot-dashed lines and exhibits a similar separation between the curves for different oxygen fugacities to that predicted by the isostress model. In contrast, the oxygen-fugacity dependence of strain rate predicted by the isostrain model, indicated by the separation between the orange and purple dashed lines, is less than that exhibited by the experimental data. Therefore, this comparison of the oxygen-fugacity dependence of the strain rate also suggests that deformation of olivine aggregates occurs closer to the isostress endmember than the isostrain endmember. Although previous work has suggested that the temperature dependence of experimental strain rates better matches those predicted by an isostrain model than those predicted by an isostress model (Hansen et al., 2016), Figure 3 demonstrates that the temperature dependences of these endmember models are sufficiently similar to not provide a robust additional discriminator.

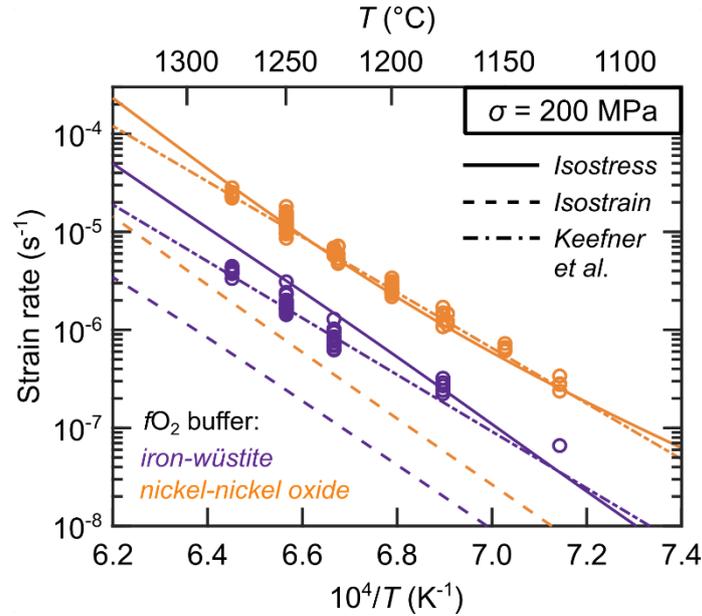

**Figure 3.** Strain rates measured in experiments and predicted by the isostress and isostrain models as a function of inverse temperature for both iron-wüstite (purple) and nickel-nickel oxide (orange) oxygen-fugacity buffers. Experimental data are normalised to a single stress of 200 MPa using a stress exponent of 3.5 (Bai et al., 1991). The flow law fitted by Keefner et al. (2011) is plotted as a dot-dashed line. Strain rates predicted by the isostress model are plotted as solid lines and those predicted by the isostrain model are plotted as dashed lines.

To explore the extent to which intragranular hardening by dislocation interactions can account for the magnitude of viscosity evolution during transient creep, we perform a similar comparison between model predictions and data collected on dunites (Keefner et al., 2011), but instead consider strain rates at the onset of deformation. The red curves in Figures 2a and 1b indicate initial strain rates predicted by a plasticity flow law describing dislocation glide at high temperatures in the absence of intragranular hardening (Breithaupt et al., 2023). The dunites typically have initial dislocation densities (Jung et al., 2020) on the order of $8\times10^{11}$ m$^{-2}$. As such, the red curves are predictions based on this dislocation density. We assume that these preexisting dislocations in the dunite cored for experiments are unlikely to coincidentally be organised and oriented in such a way that their stress fields systematically counteract the specific direction of the maximum principal stress applied during experiments, that is they generate negligible back stress during the initial increment of deformation. The predicted strain rates are one to two orders of magnitude faster than those modelled or measured at steady state. For comparison, the red data points indicate the initial strain rates measured in experiments on dunites at the onset of deformation before any hardening has occurred. Again, the initial strain rates predicted by the isostress model are in broad agreement with the data measured on dunites. The arrows in Figure 2b link data points representing the initial and steady-state strain rates during individual experiments. These arrows indicate that the transients involved reductions in strain rate by one to two orders of magnitude, similar to the difference between the isostress models that omit (red) or include (black) the effects on intragranular hardening.

In contrast to the isostress model, transients predicted by the isostrain model are inconsistent with the experimental data. Conveniently, the initial strain rates predicted by the isostrain model are equivalent to those predicted by the isostress model as no intergranular stress transfer has occurred prior to the first increment of deformation. Previous formulations of the isostrain model (Karato, 1998, 2021) have assumed that intragranular processes make a negligible contribution to transients and therefore that the isostrain model should be expressed using flow laws for each slip system calibrated at steady state. In this case, the initial strain rates would be those indicated by the solid black curves in Figures 2a and 2b and would be too slow compared to strain rates measured at the onset of deformation indicated by red data points.

Lastly, we note that the oxygen fugacity (i.e., the nickel-nickel oxide buffer) in Figure 2a is that at which the flow law for dislocation glide (Breithaupt et al., 2023) was calibrated, whereas that in Figure 2b (i.e., the iron-wüstite buffer) is approximately three orders of magnitude lower to allow comparison with several more initial strain rates measured from dunites. The agreement between predicted and measured strain rates before any hardening (red curves and data in Figure 2b), even at this lower oxygen fugacity, suggests that rates of dislocation glide may have little or no dependence on oxygen fugacity.

## 4. Discussion

*4.1. Microstructural evidence for deformation near the isostress endmember*

Our interpretation that the deformation of olivine aggregates occurs close to the isostress endmember is consistent with measurements of the variation in dislocation density with grain orientation. As dislocation density is expected to be proportional to resolved shear stress during steady-state creep (Bai and Kohlstedt, 1992; Hirth and Kohlstedt, 2015; Breithaupt et al., 2023), the isostress model predicts that grains in orientations that are calculated to have high resolved shear stress on the easy slip system will exhibit greater dislocation densities than grains in which the resolved shear stress is lower. In contrast, the isostrain model predicts that stress is transferred away from grains that are well oriented for the easy slip system and the stress is instead supported by grains in other orientations. Therefore, the isostrain model predicts that this latter population of grains should in fact support greater than average stresses and exhibit greater than average dislocation densities. Measurements of the variation in dislocation density with grain orientation demonstrate that grains that are well oriented for the easy slip system do exhibit greater dislocation densities than grains that are badly oriented for the easy slip system (Karato, 1988; Karato and Lee, 1999; Lee et al., 2002; Farla et al., 2011) and therefore indicate that deformation of olivine aggregates occurs closer to the isostress endmember than the isostrain endmember.

The same phenomenon is indicated indirectly by olivine aggregates that have been deformed by simple shear and have subsequently undergone partial static annealing by abnormal grain growth. Abnormal grain growth is a process by which a small subset of grains grow preferentially by consuming their neighbours. This process can occur by grain-boundary migration that is driven by differences in the density of dislocations and their stored energy between adjacent grains (e.g., Boneh et al., 2017). Grains with low dislocation density, and therefore less stored energy, grow at the expense of neighbouring grains with high dislocation density and more stored energy. During deformation in simple shear, grains reorient and form a CPO that results in high resolved shear stress on the easy slip system in the majority of grains (e.g., Hansen et al., 2016). During subsequent static annealing, grains with orientations that lie outside the CPO grow preferentially at the expense of grains within the CPO (Boneh et al., 2017, 2021). This observation suggests that grains with orientations within the CPO and thus high resolved shear stress on the easy slip system had greater dislocation densities and associated internal energies than those with orientations outside the CPO. This phenomenon again implies that deformation occurs closer to the isostress endmember than the isostrain endmember and, as it occurs in both experimental (Boneh et al., 2017) and natural (Boneh et al., 2021) samples, implies that deformation under natural conditions is also close to isostress.

*4.2. Intergranular strain compatibility by intragranular strain heterogeneity*

The fact that experimental strain rates are in agreement with those predicted by the isostress model indicates that strain compatibility among adjacent grains is not achieved by significant transfer of stress in a systematic manner between subpopulations of grains with different orientations. Alternatively, strain compatibility can be achieved by gradients in plastic strain near grain boundaries (Paterson, 1969), which require the presence of geometrically necessary dislocations (Ashby, 1970; Fleck et al., 1994). As geometrically necessary dislocations are ubiquitous in deformed olivine aggregates (Wallis et al., 2019, 2021, 2022), we suggest that the occurrence of the heterogeneous plastic strain that they indicate allows olivine aggregates to deform close to the isostress endmember. To investigate the generality of this behaviour, we consider the impact of each of these strain-compatibility mechanisms on the energy budget of deformation. If the difference in strain between a grain and the bulk is $\Delta\varepsilon$, then strain compatibility can be maintained by transferring a stress $\mu\Delta\varepsilon$ to or away from the grain, where $\mu$ is the shear modulus. The excess elastic energy stored by this mechanism is $(1/2)\mu(\Delta\varepsilon)^2$. Alternatively, this strain difference can be accommodated by gradients in plastic strain, resulting in a density of geometrically necessary dislocations given by $\rho_{\text{GND}} = |\Delta\varepsilon|/(4bD)$, where $b$ is the magnitude of the Burgers vector and $D$ is the grain size (Ashby, 1970). The stored energy associated with the stress fields of these dislocations (58) is $(1/2)\mu b^2 \rho_{\text{GND}} = (1/8)\mu(b/D)|\Delta\varepsilon|$. Crucially, the energy stored by dislocations is linear in the strain difference, whereas the energy stored by elastic stress transfer is quadratic in this quantity. Thus, for strain differences exceeding a critical value, $b/(4D)$, it is more energetically favourable to maintain strain compatibility by gradients in plastic strain, with associated geometrically necessary dislocations, than by further stress transfer. For the grain size of 0.9 mm used in the experiments (Keefner et al., 2011) considered in Figures 2 and 3, which is comparable to those in much of the upper mantle (e.g., Ave Lallemant et al., 1980; Behr and Hirth, 2014), the critical strain difference is approximately $10^{-7}$, which corresponds to a total transferred stress of only approximately 10 kPa. This predicted transferred stress is orders of magnitude less than differential stresses applied in the laboratory (e.g., Keefner et al., 2011) and differential stresses that are inferred to operate in regions of deforming mantle (e.g., Ave Lallemant et al., 1980; Behr and Hirth, 2014; Hansen and Warren, 2015). Consequently, systematic stress transfer among grains with different lattice orientations, which is part of the isostrain model, seems unlikely to make a significant contribution to deformation in either the laboratory or the mantle.

*4.3. Implications for steady-state dislocation creep*

To explore the implications of discriminating between the isostress and isostrain endmembers of dislocation creep for steady-state flow in the upper mantle, we analyse the ratio of viscosities predicted by each endmember model as a function of temperature and oxygen fugacity as presented in Figure 4. We note that these ratios are independent of differential stress as the stress sensitivities of each slip system are identical. Whilst the strain rates, and hence viscosities, predicted by the isostress and isostrain models differ by approximately one order of magnitude under the high temperatures and relatively low pressures of laboratory experiments (Figures 2 and 3), this difference increases to two or three orders of magnitude at the lower temperatures and higher pressures typical of the shallow upper mantle (Figure 4). For example, the difference between the viscosities predicted by each model is particularly pronounced under the low temperatures and high oxygen fugacities typical of the shallow portions of subduction zones. As the observations described above indicate that the isostress endmember provides an accurate description of the viscosity of olivine, this difference implies that the isostrain endmember (Karato, 1998, 2021; Masuti and Barbot, 2021) is an inappropriate conceptual basis for models of viscosity evolution applied to geodetic data collected in the aftermath of major subduction-zone earthquakes (e.g., Masuti et al., 2016; Muto et al., 2019).

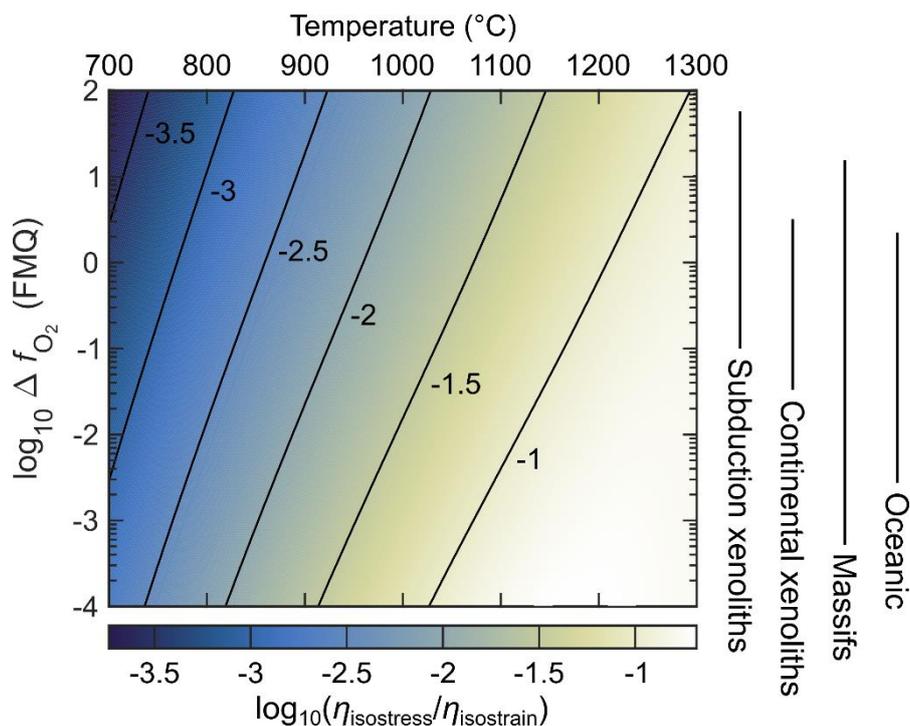

**Figure 4.** The ratio of steady-state viscosities predicted by the isostress ($\eta_{\text{isostress}}$) and isostrain ($\eta_{\text{isostrain}}$) models as a function of temperature and oxygen fugacity ($f_{O_2}$) over ranges relevant to the upper mantle. Oxygen fugacities and strain rates were calculated as functions of temperature at a pressure of 3.3 GPa corresponding to a depth of approximately 100 km. Oxygen fugacities are reported relative to that at the fayalite-magnetite-quartz (FMQ) buffer. The vertical bars on the right indicate the ranges of oxygen fugacities estimated from mantle rocks exhumed as subduction-zone xenoliths, continental xenoliths, orogenic peridotite massifs, and oceanic peridotites (Frost and McCammon, 2008).

*4.4. Implications for transient dislocation creep*

We can assess the significance of discriminating between isostress and isostrain models in the context of transient creep by comparing the predictions of more complex models that explicitly include the evolution of viscosity with strain/time due to either dislocation interactions, which are assumed to be the dominant cause of hardening in an isostress material, or grain interactions, which are assumed to dominate hardening in an isostrain material. As an example, we explore transient viscosity evolution of the upper mantle in a simplified postseismic scenario using the dislocation-interaction model of Breithaupt et al. (2023) (Section 2.4.2) and the grain-interaction model of Karato (2021) (Section 2.4.1).

To approximate postseismic deformation, we set up a simple zero-dimensional model with an initial, background stress of 0.1 MPa representing typical interseismic stress in the mantle wedge below a locked portion of a subduction interface (e.g., Sobolev and Muldashev, 2017; Govers et al., 2018). For both models, we integrate internal-stress changes until strain rate and internal stress reach the steady-state values associated with the applied macroscopic interseismic stress. Then, we impose a step increase in stress, representing the increase in stress in the upper mantle below a coseismic rupture, and keep this stress constant for the remaining time. Here, we integrate internal-stress changes until a new steady-state strain rate is reached. Whilst postseismic deformation in nature likely involves more complex stress histories, such as stress relaxation, this simplified scenario allows a transparent comparison of model behaviours for the simple purpose of highlighting similarities and significant differences.

Figure 5a presents the evolution of viscosity predicted by the models based on either dislocation interactions or grain interactions in the aftermath of two example stress changes. After the stress increase, both models predict strain rates that are higher, and hence viscosities that are lower, than those at eventual steady state. The model based on dislocation interactions predicts reductions in viscosity over timescales on the order of $10^1$–$10^2$ days, driven by increases in the number of dislocations, followed by viscosity increase over similar intervals, due to increasing back stress from dislocation interactions. In contrast, the model based on grain interactions predicts that the viscosity minimum occurs immediately after the stress change and is followed by a gradual monotonic increase in viscosity over $10^3$–$10^4$ days.

To assess the typical viscosity magnitudes of the transients, Figure 5b presents the ratio of the minimum viscosity to the steady-state viscosity over final stresses in the range 0.1–1 MPa and temperatures in the range 1250–1400°C. The viscosity ratios are fairly insensitive to final stress and temperature over these ranges and are on the order of $10^{-2}$ for the model based on dislocation interactions and $10^{-1}$ for the model based on grain interactions. The magnitudes of viscosity change of approximately two orders of magnitude predicted based on dislocation interactions are broadly similar to those inferred from postseismic geodetic time series after, for example, the 2012 $M_w$ 8.6 Indian Ocean earthquake (Masuti et al., 2016) or the 2011 $M_w$ 9.0 Tōhoku earthquake (Muto et al., 2019).

To characterise the typical durations of the transients, we define a decay time, representing the time that it takes for the viscosity to return to an intermediate reference viscosity, $\eta_{\text{ref}}$, that we define as the geometric mean of the minimum and steady state viscosities ($\eta_{\text{ref}} = \sqrt{\eta_{\text{min.}} \cdot \eta_{\text{steady-state}}}$), which is useful when viscosities change by multiple orders of magnitude. Points on the curves in Figure 5a mark this geometric mean viscosity on the viscosity-time series. Figures 5c and 5d present the decay times as functions of final stress and temperature predicted by the models based on dislocation interactions and grain interactions respectively over the same ranges as Figure 5b. Decay times predicted by the model based on dislocation interactions are typically on the order of $10^1$–$10^2$ days whereas those predicted by the model based on grain interactions span a wider range of approximately $10^1$–$10^4$ days. These decay times are broadly consistent with those implied by postseismic geodetic data, which are typically on the order of $10^2$–$10^3$ days (e.g., Freed et al., 2012, 2017; Broerse et al., 2015; Masuti et al., 2016; Muto et al., 2019; Peña et al., 2020), but do not provide a strong discriminator between the relevance of the two models.

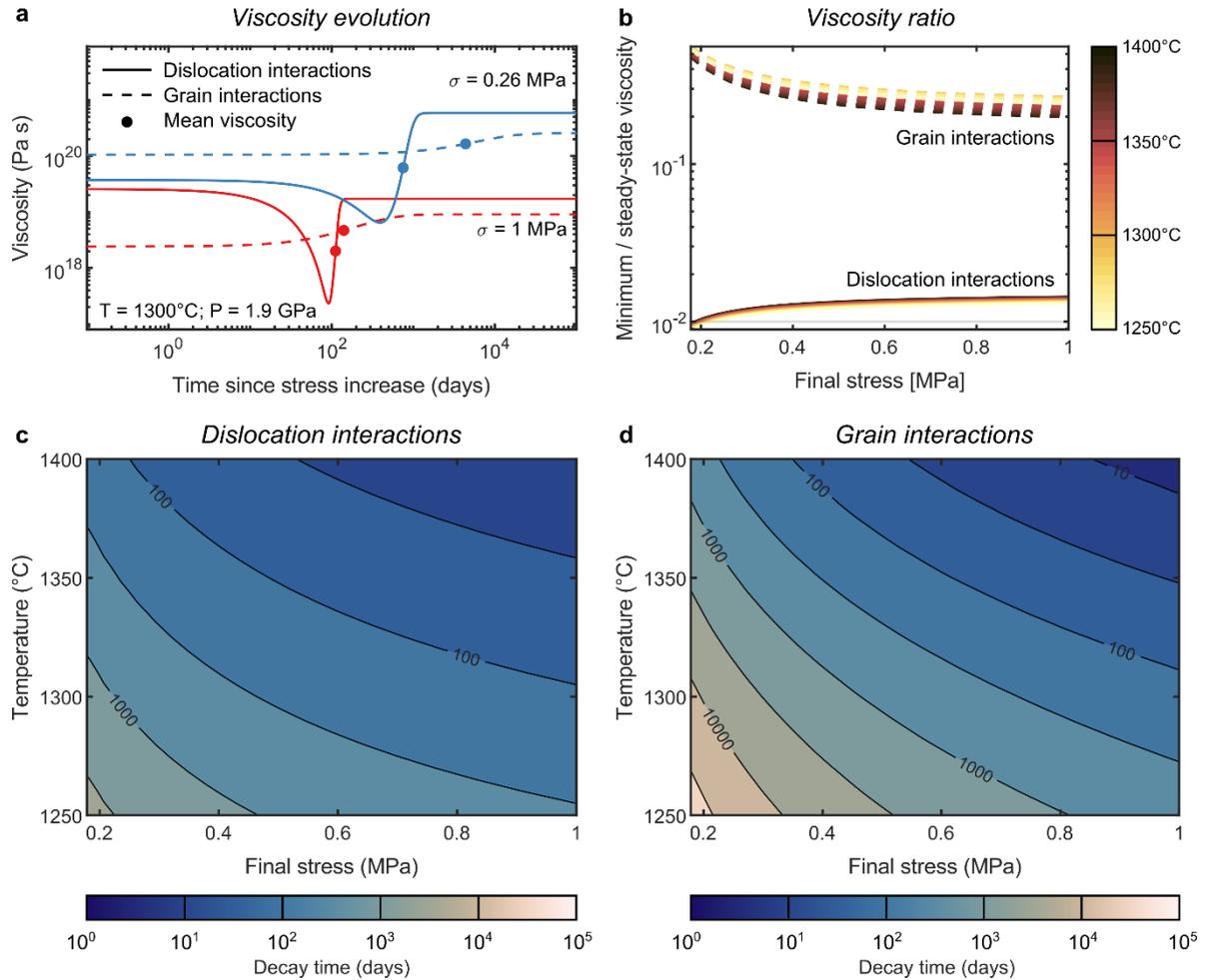

**Figure 5.** Predictions of viscosity evolution from models of dislocation interactions and grain interactions. Starting from an initial stress of 0.1 MPa, we increase the stress to a new constant stress over a range of stresses and temperatures and compute the viscosity evolution using models of dislocation interactions (Breithaupt et al., 2023) and grain interactions (Karato, 2021). In the model of grain interactions, we use flow laws for the weak (010)[100] and strong (010)[001] slip systems (Bai et al., 1991). **(a)** The evolution of viscosity for both models, at a temperature of 1300°C and a pressure corresponding to 60 km depth, for two different final stresses. The dot denotes the reference viscosity, defined as the geometric mean of the minimum and steady-state viscosities, for each curve used to determine the decay times illustrated in (c) and (d). **(b)** The ratio between minimum viscosity after the stress increase and the subsequent steady-state viscosity as a function of final stress for a range of temperatures. **(c)** The decay time that it takes for viscosity to increase to the reference viscosity in the model of dislocation interactions. **(d)** The decay time that it takes for viscosity to increase to the reference viscosity in the model of grain interactions.

These order-of-magnitude contrasts in the magnitude and duration of viscosity evolution during transient creep predicted by the models based on dislocation interactions and grain interactions demonstrates the importance of discriminating between the isostress and isostrain material behaviours with which each of these models of transient creep is respectively associated. Our finding that the rheological behaviour of olivine aggregates matches that predicted by the isostress endmember indicates that transient creep occurs primarily due to dislocation interactions, rather than the grain interactions associated with the isostrain endmember. The analysis in Figure 5 suggests that, under conditions typical of the upper mantle, the transients arising from dislocation interactions involve greater viscosity changes, are briefer in duration, and are more complex in form than those predicted by the isostrain model that forms the conceptual basis of recent analyses of postseismic geodetic data (Masuti et al., 2016, Muto et al., 2019). Overall, this simplified example demonstrates the importance of our results in discriminating the most appropriate model to employ in analyses of postseismic deformation as the different models predict significantly different rheological behaviours.

Lastly, we note that several previous works have represented transient creep of olivine by fitting phenomenological models, such as Burgers models, to experimental data (Post, 1977; Smith and Carpenter, 1987; Hanson and Spetzler, 1994; Chopra, 1997; Masuti and Barbot, 2021; Ohuchi et al., 2024). Such fitting exercises may generate equations that superficially reproduce some aspects of the behaviours observed under the conditions of a particular set of experiments. However, even if a close fit is obtained, two important points must be borne in mind in relation to the applicability of such models to analysis of deformation in Earth's interior. First, the lack of a robust microphysical basis to these models means that extrapolation beyond their calibration range does not necessarily yield accurate predictions of deformation behaviours. Rather, the interactions among underlying physical processes (e.g., rates of dislocation glide and climb) may play out differently in Earth's interior due to their different dependencies on stress, temperature, and other factors, potentially leading to significantly different behaviours (e.g., Paterson, 1976). As such, we reiterate the long-standing point that the development of models that capture the important microphysics should take precedence over cursory fitting of phenomenological equations (e.g., Paterson, 1976). Second, although the concept of grain interactions has been linked to Burgers models as a loose, post-hoc, physical basis (Karato, 1998, 2021; Masuti and Barbot, 2021; Ohuchi et al., 2024), the finding that deformation of olivine aggregates occurs at the isostress endmember during transient and steady-state dislocation creep in laboratory experiments (Section 3), along with the microstructural and energy arguments that this is also the case in the upper mantle (Sections 4.1 and 4.2), mean that the concept of grain interactions can no longer be appealed to. Instead, modelling transient creep of olivine as occurring due to dislocation interactions (Hansen et al., 2021; Breithaupt et al., 2023) in an isostress material (Figures 2 and 3) provides a more robust microphysical approach.

## 5. Conclusions

Our comparison of the steady-state strain rates measured on single crystals of olivine to those measured on aggregates, properly accounting for the effects of grain orientation (Figure 1), oxygen fugacity, and pressure, demonstrates that aggregates deform at strain rates that are in excellent agreement with those predicted by the isostress model but about an order of magnitude faster than those predicted by the isostrain model (Figures 2 and 3). Unlike the isostrain model, the isostress model also well predicts the magnitude of strain hardening observed in laboratory experiments (Figure 2). When extrapolated to conditions typical of the upper mantle, the steady-state strain rates predicted by the isostress and isostrain models (Figure 4), along with the magnitudes, durations, and forms of transients resulting from associated microphysical processes (Figure 5), all differ by orders of magnitude. These results challenge theoretical work (Karato, 1998, 2021; Masuti et al., 2019; Masuti and Barbot, 2021) and recent analyses of geodetic data (Masuti et al., 2016; Muto et al., 2019) that either explicitly or implicitly assume *a priori* that olivine aggregates deform close to the isostrain endmember with transients arising from stress transfer between soft and hard slip systems in grains of different orientation. Instead, laboratory data demonstrate that olivine aggregates deform close to the isostress endmember, which we suggest is possible primarily due to the occurrence of heterogeneous intragranular strain fields associated with geometrically necessary dislocations. As such, recent models for transient and steady-state plasticity and dislocation creep of olivine aggregates (Hansen et al., 2019; 2021; Breithaupt et al., 2023), in which viscosity is controlled by the density of dislocations and the strength of interactions among them, provide a more robust physical basis for analysing the rheological behaviour of the upper mantle.

**Acknowledgements**

The authors thank Lars Hansen for helpful conversations and David Kohlstedt for supplying the sample of Åheim dunite. The authors acknowledge funding from the Netherlands Organisation for Scientific Research, User Support Programme Space Research grant ALWGO.2018.038. to David Wallis and Taco Broerse and grant ENW.GO.001.005 to Taco Broerse, David Wallis, and Thomas Breithaupt. David Wallis acknowledges funding from a UK Research and Innovation Future Leaders Fellowship MR/V021788/1. Thomas Breithaupt acknowledges funding from a Research Fellowship from the Royal Commission for the Exhibition of 1851.

**Data and software availability:**

All data and code developed will be made available through a Zenodo deposit upon acceptance and is available during review upon request.

# Supporting information for

**Transient and steady-state dislocation creep of olivine controlled by dislocation interactions at the isostress endmember**

David Wallis,[1]* Thomas Breithaupt,[1] Taco Broerse[2]

[1]Department of Earth Sciences, University of Cambridge, Downing Street, Cambridge, CB2 3EQ, UK.

[2]Department of Geoscience and Remote Sensing, Delft University of Technology, Delft, 2628 CN, The Netherlands.

*dw584@cam.ac.uk

This supplementary material provides the parameter values used in the models of grain interactions and dislocation interactions in Table S1 and Table S2, respectively. Values in Table S1 are from Bai et al. (1991), their Table 4 (orthopyroxene silica buffer), converted to SI units. Values in Table S2 are from Breithaupt et al. (2023) and are inferred values for the full model where the dislocation density was eliminated.

**Table S1.** Parameters used in the model of grain interactions.

| Parameter | Name | Values | Unit |
|---|---|---|---|
| $A_a$ (010)[001]<br>$A_b$ | Pre-exponent at steady state | $1.6677 \cdot 10^{-17}$<br>$3.6702 \cdot 10^{-17}$ | s$^{-1}$Pa$^{-n}$Pa$^{-m}$ |
| $m_a$ (010)[001]<br>$m_b$ | Oxygen fugacity exponent | 0.02<br>0.23 | - |
| $Q_a Q_b$ (010)[001] | Reference activation energy | $5.4 \cdot 10^5$ | J/mol |
| $n_a n_b$ (010)[001] | Stress exponent | 3.5 | - |
| $A_c$ (010)[100]<br>$A_d$<br>$A_e$ | Pre-exponent at steady state | $3.1548 \cdot 10^{-25}$ $4.1056$<br>$2.1297 \cdot 10^{-22}$ | s$^{-1}$Pa$^{-n}$Pa$^{-m}$ |
| $m_c$ (010)[100]<br>$m_d$<br>$m_e$ | Oxygen fugacity exponent | 0.36<br>0.10<br>0.15 | - |
| $Q_c$ (010)[100]<br>$Q_d$<br>$Q_e$ | Reference activation energy | $2.3 \cdot 10^5$<br>$10^7$<br>$2.9 \cdot 10^5$ | J·mol$^{-1}$ |
| $n_c n_d n_e$ (010)[100] | Stress exponent | 3.5 | - |
| $P$ | Pressure | $1.9424 \cdot 10^9$ | Pa |
| $\Delta V$ (010)[100] | Activation volume | $1.2 \cdot 10^{-5}$ | m$^3$·mol$^{-1}$ |
| $\Delta V$ (010)[001] | Activation volume | $3 \cdot 10^{-6}$ | m$^3$·mol$^{-1}$ |
| $\mu$ | Elastic modulus | $6.5 \cdot 10^{10}$ | Pa |

**Table S2.** Parameters used in the model of dislocation interactions.

| Parameter | Name | Value | Unit |
|---|---|---|---|
| $A$ | Pre-exponent | $8.7096 \cdot 10^{-6}$ | s⁻¹Pa⁻² |
| $b$ | Magnitude of Burgers vector | $5 \cdot 10^{-10}$ | m |
| $\alpha$ | Taylor constant | $2.46$ | - |
| $\beta$ | Threshold-stress constant | $2$ | - |
| $d$ | Grain size | $10^{-3}$ | m |
| $M$ | Taylor-stress modulus | $1.35 \cdot 10^{11}$ | Pa |
| $\mu$ | Elastic modulus | $6.5 \cdot 10^{10}$ | Pa |
| $Q$ | Reference activation energy | $4.5 \cdot 10^{5}$ | J·mol⁻¹ |
| $\Delta V$ | Activation volume | $1.5 \cdot 10^{-5}$ | m³·mol⁻¹ |
| $\mathcal{R}^*_{\text{gb}}$ | Rate coefficient for grain boundary diffusion | $3.3884 \cdot 10^{-21}$ | s⁻¹Pa⁻⁴ |
| $\mathcal{R}^*_{\text{pipe}}$ | Rate coefficient for pipe diffusion | $1.122 \cdot 10^{-31}$ | s⁻¹Pa⁻⁵ |
| $\sigma^*_{\text{P}}$ | Peierls stress | $3.1 \cdot 10^{9}$ | Pa |